\documentclass{ecai} 

\usepackage{latexsym}
\usepackage{amssymb}
\usepackage{enumitem}
\usepackage{times}
\usepackage{soul}
\usepackage{url}
\usepackage[small]{caption}
\usepackage{graphicx}
\usepackage{amsmath}
\usepackage{amsthm}
\usepackage{booktabs}
\usepackage{algorithm}
\usepackage{multirow}
\usepackage{algorithmic}
\usepackage{xcolor}
\usepackage{balance}
\usepackage{stfloats}   

\usepackage{xspace}

\newcommand{\ie}{\emph{i.e.}\xspace}

\newcommand{\ourmethod}{Nissist~}

\newcommand{\BibTeX}{B\kern-.05em{\sc i\kern-.025em b}\kern-.08em\TeX}

\begin{document}


\begin{frontmatter}


\paperid{36} 


\title{%
\raisebox{-0.3cm}{\includegraphics[width=1cm, height=1cm]{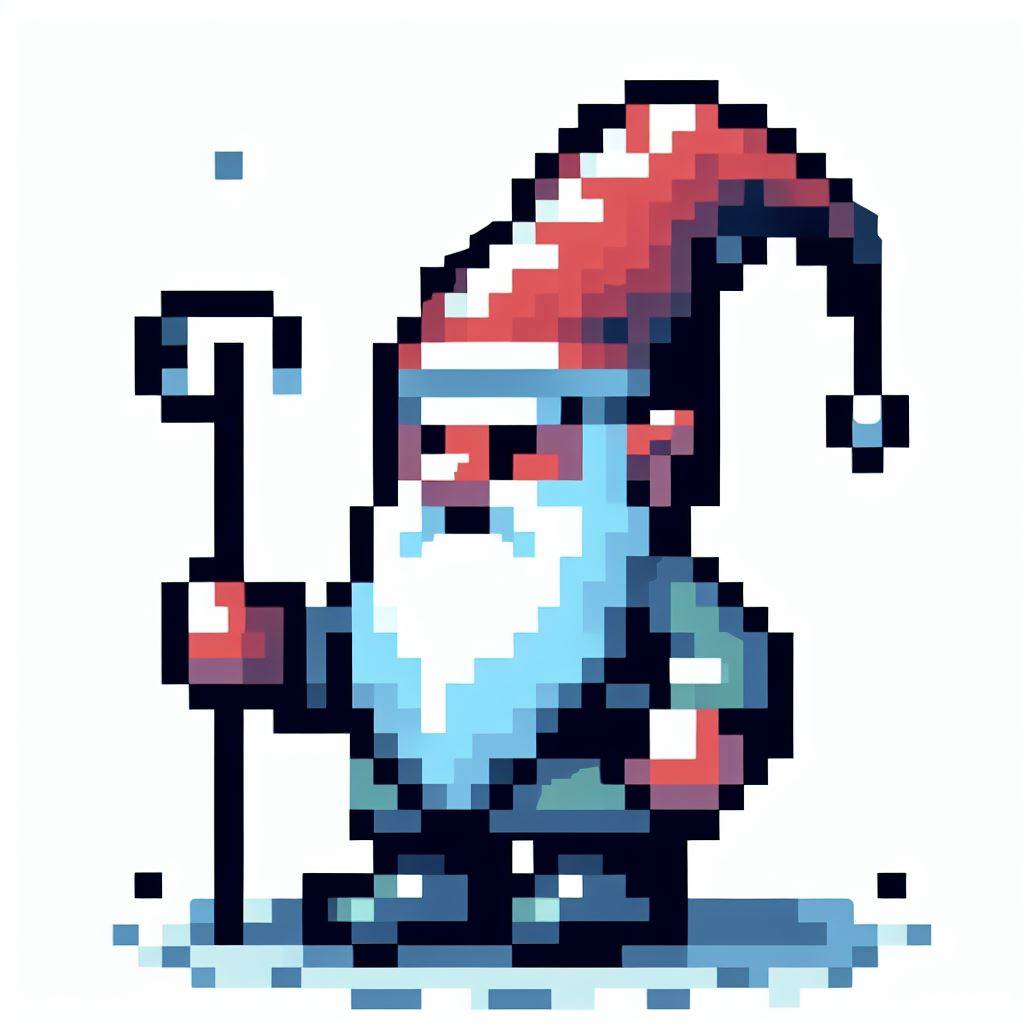}}\ {Nissist: An Incident Mitigation Copilot based on Troubleshooting Guides}
}


\renewcommand*{\thefootnote}{\fnsymbol{footnote}}
\newcommand{\sharedthanks}{\footnotemark[*]}

{\author[A]
{\fnms{Kaikai}~\snm{An}\thanks{This work is done during the internship at Microsoft.}}
\author[B]
{\fnms{Fangkai}~\snm{Yang}}
\author[A]
{\fnms{Junting}~\snm{Lu}\sharedthanks}
\author[B]
{\fnms{Liqun}~\snm{Li}}
\author[B]
{\fnms{Zhixing}~\snm{Ren}}
\author[B]
{\fnms{Hao}~\snm{Huang}}
\author[B]
{\fnms{Lu}~\snm{Wang}}
\author[B]
{\fnms{Pu}~\snm{Zhao}}
\author[B]
{\fnms{Lingling}~\snm{Zheng}}
\author[B]
{\fnms{Yu}~\snm{Kang}}
\author[B]
{\fnms{Hua}~\snm{Ding}}
\author[B]
{\fnms{Qingwei}~\snm{Lin}}
\author[B]
{\fnms{Saravan}~\snm{Rajmohan}}
\author[B]
{\fnms{Dongmei}~\snm{Zhang}}
\author[B]
{\fnms{Qi}~\snm{Zhang}}
}

\address[A]{Peking University}
\address[B]{Microsoft}

\renewcommand*{\thefootnote}{\arabic{footnote}}

\begin{abstract}
Effective incident management is pivotal for the smooth operation of Microsoft cloud services. In order to expedite incident mitigation, service teams gather troubleshooting knowledge into Troubleshooting Guides (TSGs) accessible to On-Call Engineers (OCEs). While automated pipelines are enabled to resolve the most frequent and easy incidents, there still exist complex incidents that require OCEs' intervention. In addition, TSGs are often unstructured and incomplete, which requires manual interpretation by OCEs, leading to on-call fatigue and decreased productivity, especially among new-hire OCEs. In this work, we propose \textit{\ourmethod}which leverages unstructured TSGs and incident mitigation history to provide proactive incident mitigation suggestions, reducing human intervention. Leveraging Large Language Models (LLM), \ourmethod extracts knowledge from unstructured TSGs and incident mitigation history, forming a comprehensive knowledge base. Its multi-agent system design enhances proficiency in precisely discerning OCE intents, retrieving relevant information, and delivering systematic plans consecutively. Through our user experiments, we demonstrate that \ourmethod significantly reduce Time to Mitigate (TTM) in incident mitigation, alleviating operational burdens on OCEs and improving service reliability. Our demo is available at \url{https://aka.ms/nissist_demo}\footnotemark[1].

\end{abstract}

\end{frontmatter}

\footnotetext[1]{Work in progress, code will be released later.}

\section{Introduction}
In the rapidly evolving landscape of cloud operation, incident management stands as a pivotal challenge for enterprise-level cloud service providers~\cite{liu2023incident,chen2020towards,sarkar2011automated} such as Microsoft, Google, and Amazon. The profound impact of incidents, exemplified by notable events such as the Amazon outage~\cite{Amazon}, underscores the need for a robust incident management system. Incidents can range from minor operational interruptions to severe system failures, with potential consequences including financial loss, operational disruption, reputational harm, and legal complications. Swiftly identifying, troubleshooting, and resolving system incidents is essential for maintaining service reliability and operational continuity~\cite{li2022intelligent,li2021fighting}.
While automated pipelines can handle low-severity incidents due to their simplicity and commonality, high-severity incidents require immediate and hands-on intervention by On-Call Engineers~(OCEs), beyond the capabilities of automated systems.
Service teams address this challenge by documenting frequent troubleshooting steps in Troubleshooting Guides (TSGs), empowering OCEs to efficiently resolve incidents~\cite{jiang2020mitigate,ghosh2022fight}. 

To investigate the effect of TSGs on incident mitigation, we analyze around 1000 high-severity incidents in the recent twelve months that demand immediate intervention from OCEs. 
Consistent with findings from prior studies~\cite{jiang2020mitigate,10.1145/3540250.3558958,jiang2023xpert}, which demonstrate the efficacy of TSGs in incident mitigation. We found that incidents paired with TSGs exhibit a 60\% shorter average time-to-mitigate (TTM) compared to those without TSGs, emphasizing the pivotal role played by TSGs. This trend is consistent across various companies, as evidenced by research~\cite{lotufo2015modelling,li2018learning}, even among those employing different forms of TSGs.
However, despite their utility, as highlighted by \cite{10.1145/3540250.3558958,chen2023empowering}, the unstructured format, varying quantity, and propensity for internal use purpose of TSGs, impede their optimal utilization. Particularly, such unstructured TSGs pose challenges for new hires and contribute to the complexity of the incident mitigation process, especially in scenarios requiring coordination across multiple teams. In addition, some TSGs are outdated, lacking the most recent knowledge of incident mitigation. The incident mitigation history provided by OCEs in the internal incident management platform serves as another valuable resource for extracting incident mitigation knowledge.

Recent works have focused on leveraging TSGs to facilitate incident mitigation process. \cite{10.1145/3540250.3558958} fine-tunes models to extract knowledge from TSGs, while \cite{jiang2023xpert,chen2023empowering} identify relevant TSGs in root cause analysis. However, the prevalent unstructured nature of existing TSGs limits the effectiveness of fine-tuning procedures, and the complexity of high-severity incidents still require human interventions. 
In this work, we propose \textit{Nissist}, aiming to reduce OCE workload and assist incident mitigation processes. Firstly, we establish a set of rigorous TSG criteria to convert unstructured TSGs into structured, high-quality formats leveraging Large Language Models~(LLMs), while also providing guidelines for OCEs when documenting new TSGs. Subsequently, we propose a novel structure of knowledge base comprising discrete executable nodes extracted from TSGs and incident mitigation history. 
Moreover, we introduce an advanced multi-agent system~\cite{wu2023autogen} designed to proficiently interpret queries, retrieve relevant knowledge nodes, and formulate actionable plans in a semi-automated manner. By interacting with Nissist, the incident mitigation trajectories are optimized, allowing OCEs to focus on challenging mitigation steps not covered by TSGs and mitigation history, thus significantly reducing direct human intervention.

\section{System Overview}
\begin{figure*}[t]
    \centering
    \includegraphics[width=0.75\textwidth]{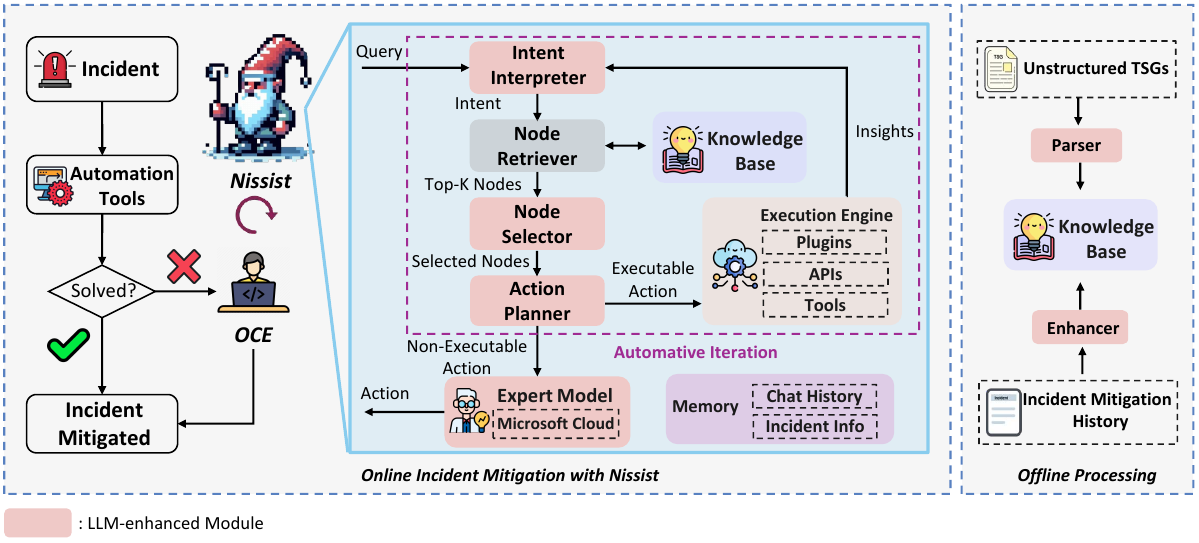}
    \caption{
    The Semi-Automated Incident Mitigation Framework with Nissist. When incidents exceed automation capabilities, OCEs engage in iterative interactions with Nissist. Nissist interprets OCE intents, retrieves knowledge from the knowledge base, and formulates actions. Executable action is conducted with the execution engine, generating insights for the next round node retrieval and action planning in an automative iteration manner (purple dashed box).  Actions that cannot be carried out by the execution engines are then delegated to OCEs for manual execution. The knowledge base is built offline with LLM-extracted knowledge from unstructured TSGs and mitigation history.}
    \label{fig:framework}
    \vspace{-3mm}
\end{figure*}

As illustrated in Figure~\ref{fig:framework}, OCEs engage in iterative interactions with Nissist when automation tools are insufficient to address an incident. Initially, Nissist constructs \textit{Knowledge Base} offline by parsing knowledge from unstructured TSGs and enhances it with knowledge from incident mitigation history not covered in TSGs. Subsequently, it iteratively processes OCE queries. Nissist is designed to mitigate incidents in a fully automated manner. However, due to the complex nature of incidents, not all actions suggested by Nissist can be automatically executed due to the lack of related execution functions in the execution engine. Thus, the non-executable actions are delegated to OCEs for manual execution, while those executale actions are passed to the execution engine, generating insights to trigger the next step of mitigation, automating the mitigation iteration (purple dashed box in Figure~\ref{fig:framework}).
Each module, powered by LLMs, serves as an agent responsible for specific tasks, including interpreting intents, selecting the most relevant knowledge (nodes from knowledge base), suggesting actions, etc. These modules concurrently communicate with each other to efficiently mitigate incidents.

\subsection{Constructing the Knowledge Base}
The primary source of knowledge is derived from unstructured TSGs, which typically encompass information on investigating and mitigating incidents. However, their unstructured nature poses challenges for traditional data retrieval methods, as the appropriate action may not always exhibit semantic or lexical similarity to the query due to unstructured nature. Additionally, incident mitigation often requires a sequential steps of actions. Chunking TSGs in retrieval methods can result in the disruption of this sequence, particularly when certain steps are spread across multiple TSGs and collectively represent the entire flow of steps.
To address this, we develop quality criteria and use LLMs to reformat original TSGs into structured ones, including background, terminology, FAQ (frequently asked questions), flow, and appendix. In particular, ``flow'' represents sequential steps of actions. We then construct a \textit{knowledge base} comprising knowledge \textit{nodes}. Each node is formated in JSON consisting of type, intent, action, and linker. ``Intent'' describes the purpose of the node, and it is used as the indexing context~\cite{douze2024faiss} which is later used to retrieve the entire node. ``Linker'' connects the outcomes of taking current actions to the intents of next step nodes. 

Parsing the structured TSGs with LLMs yields this knowledge base that facilitates easy retrieval of relevant nodes for actionable plans.
Additionally, the node-level granularity enables discovery of new connections among TSGs, including cross-team mitigation flows not presented in any raw TSGs (refer to Cross-TSG case in Section~\ref{sec:usecase}). Microsoft maintains an incident management platform where OCEs can document and collaborate during incident mitigation. Beyond unstructured TSGs, incident mitigation history serves as an additional source of knowledge. To complement any outdated or missing steps in TSGs, we have designed an enhancer that captures the latest solutions from these discussions on mitigation history.

\subsection{Muti-Agent System}
\noindent{\textbf{Intent Interpreter.}} 
This module is crucial for understanding OCE's intent and determining whether Nissist's intervention is necessary during the conversation. It guides OCEs to incident troubleshooting topics and helps refine and clarify their input. If needed, it seeks confirmation from OCEs for intent clarification. 

\noindent{\textbf{Node Retriever \& Selector.}} 
The node retriever module retrieves relevant nodes from the knowledge base concerning the clarified intent. Unlike traditional retrievers~\cite{gao2023retrieval}, which index documents or chunks, we use the clarified intent as the query to retrieve top-k nodes by comparing with indexed ``Intent'' of each node. To enhance fault tolerance, the node selector selects the most relevant ones from retrieved top-k nodes.
This is crucial as semantic discrepancies may exist despite specific matched keywords. By sourcing information from multiple nodes, Nissist enriches the knowledge context for subsequent actions. If no relevant node is found, it indicates current incident exceeds Nissist's scope and informs OCEs for intervention.

\noindent{\textbf{Action Planner.}}
This module serves as the central and critical component within Nissist, recommending appropriate actions based on the selected nodes and memory. Unlike SOTA LLM planners that automate reasoning, actions, and observations in an interleaved manner~\cite{yao2023react,valmeekam2023planning,guan2023leveraging}, our domain finds this planning style unsuitable. Fully automated plugins or tools cannot handle all incidents due to their complexity, risking wrong plugin invocation or omission of execution steps by the planning module. Hence, a semi-automated mitigation process occasionally requiring human intervention is necessary for the security purpose. Action planner generates steps based on incident complexity and plugin availability. For incidents can be covered in TSGs, exhaustive step-by-step and manual planning is unnecessary. Instead, Nissist suggests sequential steps automatable by execution engine, thereby bypassing the need for OCE input. The execution engine should be able to execute the action and analyze the execution outcomes in order to give insights. The execution engine could be available plugins, APIs or LLM-based code generator~\cite{qiao2023taskweaver}.

\noindent{\textbf{Post Processor.}}
In open-domain planning scenarios, self-reflection of planned strategies against actual observations with LLMs is practical and efficient~\cite{shinn2023reflexion,yao2022react}. However, its utility is somewhat limited in incident mitigation due to the potential for hallucinations resulting from the lack of domain-specific knowledge in pre-trained LLMs~\cite{wang2023empower,wang2023survey}, such as GPT-4~\cite{achiam2023gpt}.
To address this limitation, we integrate a pre-trained LLaMA2 model~\cite{touvron2023llama} as the expert model that has undergone supervised fine-tuning (SFT) on a corpus of Microsoft Cloud documentation~\cite{wang2023empower}. 
The expert model enhances the post-processing procedure by providing informed corrections rooted in cloud domain expertise.
\section{User Experiments and Case Study}\label{sec:usecase}
\begin{figure*}[ht]
    \centering
    \includegraphics[width=\linewidth]{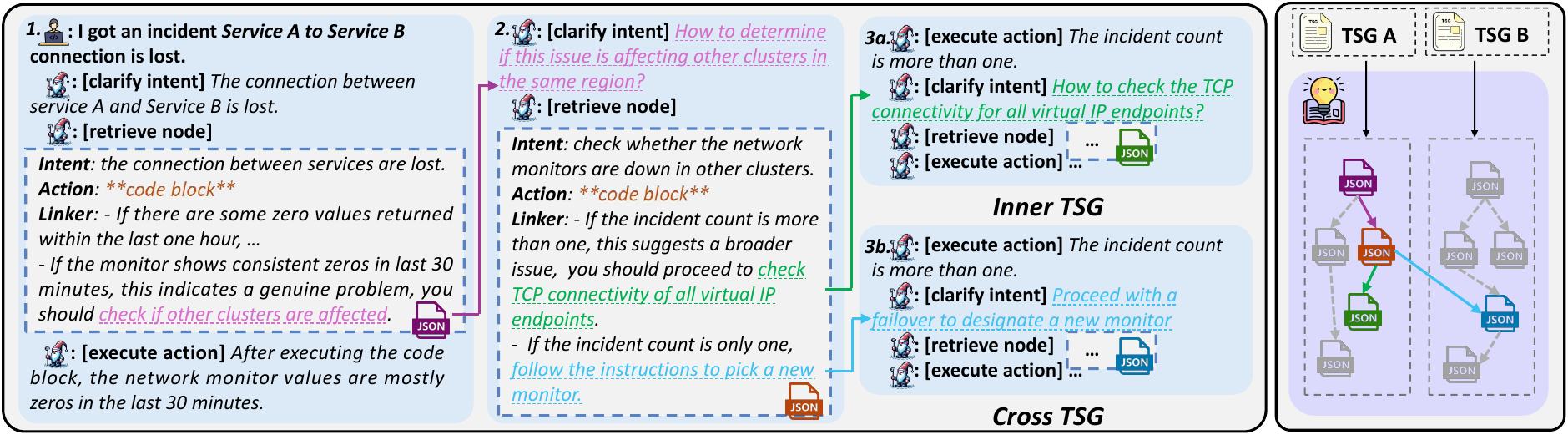}
    \caption{A use case demonstrates that Nissist mitigates the connection lost incident between \textit{Service A} and \textit{Service B}. For simplicity, only the first three iterations are presented. \textit{$3a$} \& \textit{$3b$} show two different mitigate paths due to two different execution results. In particular, \textit{$3b$} indicates that Nissist can leverage knowledge cross TSG (the blue-colored node is extracted from another TSG).}
    \label{fig:user_case}
    \vspace{-3mm}
\end{figure*}

To validate the advantages of Nissist, we conducted a human evaluation involving twenty OCEs\footnote{Varying across new-hire and experienced OCEs.} and five\footnote{Controlling time to each experiment around 30 mins to 1 hour per OCE ensures consistency, engagement, and ethical treatment.} incidents. These incidents are categorized as simple or hard based on their mitigation history. Note that simple incidents here are not easy ones that can be handled with automated mitigation tools. Each OCE is tasked with mitigating all five incidents. To ensure a fair comparison, only one mitigation approach is assigned to each incident, either with Nissist or manual mitigation. We distribute the incidents to ensure each incident gets an equal number of Nissist mitigation and manual mitigation. We demonstrate the effectiveness of Nissist and report several metrics including Success Rate (SR): whether the incident can be mitigated without human intervention; Human Intervention (IR): steps need human intervention; Turns: numebr of mitigation turns; TTM reduction (TTM $\downarrow$) compared with manual mitigation.

\begin{table}[h]
    \small
    \caption{User experiment results on metrics.}
    \centering
    \begin{tabular}{lcccc}
    \toprule
    Category & SR  & IR & Turns & TTM $\downarrow$ \\
    \midrule
    Simple & 77.19\%  & 11.28\% & 2.56 & 98.93\% \\
    Hard & 52.63\% & 15.79\% & 5.74 & 94.85\% \\
    \bottomrule
    \end{tabular}
    \label{tab:experiment}
    \vspace{-2mm}
\end{table}

Table~\ref{tab:experiment} shows a significant improvement in TTM reduction compared to manual mitigation which either requires substantial mitigation experience or involve navigating through unstructured TSGs. Specifically, Nissist achieves a TTM reduction of 98.93\% for simple incidents and 94.85\% for hard ones. Hard incidents requires additional turns for mitigation due to their complex nature. Nissist shows a 77.19\% full automation SR for simple incidents and 52.63\% SR for hard ones, demonstrating a notable reduction of manual efforts. With some minor human intervention (11.28\% for simple incidents and  15.79\% for hard ones), incidents can be effectively mitigated. 

Figure~\ref{fig:user_case} illustrates a use case demonstrating industrial practices with Nissist. Given an OCE query ``\textit{Service A to Service B connection is lost}'', Nissist interprets the query, identifies the intent, and uses it to retrieve and select the most relevant node. The action planner then fills the given parameters, such as service information, into the code block, \ie, a Kusto query with parameter placeholders in this use case. After the action is passed to the execution engine, the outcome indicates ``\textit{the network monitor values are mostly zeros in the last 30 mins}''. Nissist correlates this outcome with the ``Linker'' in the retrieved node, where this outcome indicates ``\textit{a genuine problem and should check if other clusters are affected}''. Then Nissist generates a new intent ``\textit{How to determine if this issue is affecting other clusters}'' for the next round of interaction automatically. This interaction continues until the incident is mitigated or requires human intervention. Additionally, Nissist digests all TSGs into knowledge base, making it possible to discover connections between nodes located in different TSGs. For example, \textit{$3b$} in Figure~\ref{fig:user_case} demonstrates another execution result which requires knowledge from a different TSG. Previously, it requires OCEs to take great efforts searching for the knowledge in other TSGs, which often did not list such knowledge in their titles. 
\section{Conclusion}
We address the incident mitigation challenges in Microsoft by optimizing TSG usage and reducing human effort. We introduce Nissist, which leverages LLMs to digest TSGs and incident mitigation history into a knowledge database. We establish a multi-agent system with semi-automation to precisely detect OCE intents, retrieve relevant nodes, and provide stepwise actions. Our experiments show reduced TTM and significantly alleviated OCE workload.


\newpage
\bibliography{ecai24}

\end{document}